\documentclass[useAMS,usenatbib]{mn2e}
\usepackage{graphicx,amsmath,multirow,amssymb}
\usepackage{natbib}
\usepackage{graphicx}
\usepackage{epstopdf}
\usepackage{color}

\def\cm2{cm$^{-2}$}

\newcommand{\II}{\,{\sc ii}} 
\newcommand{\comment}[1]{}

\def\simgt{\lower.5ex\hbox{$\; \buildrel > \over \sim \;$}}
\def\simlt{\lower.5ex\hbox{$\; \buildrel < \over \sim \;$}}

\title[V381 Lac]{The variable  V381 Lac and its possible connection with the R\,CrB phenomenon  }
\author[Rossi et al.]{ C. Rossi$^{1}$, F. Dell'Agli$^{1}$, A. Di Paola$^{2}$,  K.S. Gigoyan$^{3}$, 
R. Nesci$^{4}$ \\
$^1$Dipartimento di Fisica, Universit\`a di Roma ``La Sapienza'', P.le Aldo Moro 5, 00185, Roma, Italy \\
$^2$INAF -- Osservatorio Astronomico di Roma, Via Frascati 33, 00040, Monte Porzio Catone (RM), Italy \\
$^{3}$V. A. Ambartsumian Byurakan Astrophysical Observatory (BAO) and Isaac Newton Institute of Chile, \\
Armenian Branch, Byurakan 0213, Aragatzotn province, Armenia\\
$^{4}$INAF/IAPS, via Fosso del Cavaliere 100, 00133 Roma, Italy\\
\thanks  {Based on observations collected with the Cassini Telescope at Loiano station of the  INAF$-$Bologna Astronomical Observatory,   and with the Copernico Telescope of the INAF$-$Padova Astronomical Observatory  }
}
\begin{document}
\date{Accepted 2015 xx xx., Received 2015 xx xx; in original form 2015 xx xx}
\pagerange{\pageref{firstpage}--\pageref{lastpage}} \pubyear{2015}
\maketitle

\label{firstpage}

\begin{abstract}

We have performed new medium resolution spectroscopy, optical and near infrared photometry to monitor the variability of the AGB carbon star V 381 Lac.
Our observations revealed rapid and deep changes in the  spectrum and extreme  variability in the optical and near infrared bands. Most notably we observed the change of  NaI D lines  from deep absorption to emission,  and the progressive growing of the [N{\II}] doublet 6548-6584\AA ~emission,  strongly related to the simultaneous  photometric fading. 
V381 Lac occupies regions of 2MASS and WISE colour-colour diagrams typical of stars with dust formation in the envelope. The general framework emerging from the observations of V381 Lac is that of a cool AGB  carbon star undergoing episodes of  high mass ejection and severe occultation of the stellar photosphere reminiscent of those characterising  the RCB phenomenon.

Comparing the Spectral Energy Distribution  obtained with the theoretical model for AGB evolution with dust in the circumstellar envelope, we can identify V381 Lac as the descendant of a star of initial mass $\sim $2M$_{\odot}$, in the final AGB phases, evolved into a carbon star by repeated Third Dredge Up episodes. According to our model
the star is moderately obscured ($\tau_{10} \sim 0.22$) by dust, mainly formed by amorphous carbon ($\sim 80 \%$) and SiC ($\sim 20\%$), with dust grain dimensions around $\sim 0.2 \mu m$ and $0.08 \mu m$ respectively.
 \end{abstract}


\section{INTRODUCTION}
  Among the AGB stars, a small fraction of carbon rich stars show  an extreme variability reminiscent of the hotter R\,CrB supergiants,  e.g. \citep{feast03, white06}.   Erratic optical variability with large amplitude and dramatic spectral changes are the main signatures of these stars: the fast  decline  of several magnitudes in luminosity is interpreted as caused by enhanced mass loss events.
The very extended atmospheres produce the ejection of clouds (puffs) where carbon rich dust condense in grains. The extinction events are attributed to these dust clouds in our line of sight: either a dust cloud is passing between the photosphere and us, or a  sudden very strong ejection of dust by the star itself takes place.
 Colour changes also occur, being the object redder when fainter.
The photometric changes have  spectacular spectroscopic  fallout: molecular and atomic lines develop in emission during the fading  to turn back in absorption when the star recovers the bright state  \citep{white06, clayton96}.

Besides the optical variations, the infrared excess is a well known characteristics of the RCB and of their cooler counterpart, the DY Per  stars, showing  a spectrum more similar to  that of an ordinary N AGB type supergiant \citep[see e.g.][]{feast97, clayton96}. 
The advent of several infrared satellites improved the  wavelength coverage of the energy distribution and stimulated the use of mid infrared colour-colour diagrams to select new candidates,  \citep{tisse12, tisse13, miller12, lee15}.
 Differences and similarities have been extensively  discussed and categorized  although the behaviour of  strongly variable AGB stars makes them  difficult  to be identified  unless located in well monitored fields, while scattered observations can yield to misclassification of possibly interesting objects. 
 
 The case of V381\,Lac (also known as FBS 2213+421) is paradigmatic.
Over the years the star has been subject to periods of  high luminosity  and of  rapid, severe obscuration episodes.  
 Classified  as a Dwarf Nova by \citep{Dahlm96} and by \cite{hoard02} on the basis of the photometric variability, 
 then as an M5-M6 star by  \citet{gig06} from a low dispersion objective-prism plate of the First Byurakan Survey  \citep{Mark89},  it was finally recognised as a Carbon star by \cite{gig09}. 
To clarify its nature we started since 2012 a photometric and spectroscopic campaign and collected the available data from literature and public catalogs. This paper is organised inthe following way:
in Section \ref{Obs}  we describe the observational material. In Section \ref{DA} we analyse  the flux variability, the spectroscopic changes and the position of V381 Lac in several infrared colour-colour plots.
In Section \ref{theo} we present a possible interpretation of the Spectral Energy Distribution on the basis of our evolutionary models for dusty carbon rich stars.
In Section \ref{Concl} we discuss the last event of strong flux variation  and  draw our conclusions.

%
\begin{figure}
\centering
\includegraphics[width=6cm,height=6cm,angle=0]{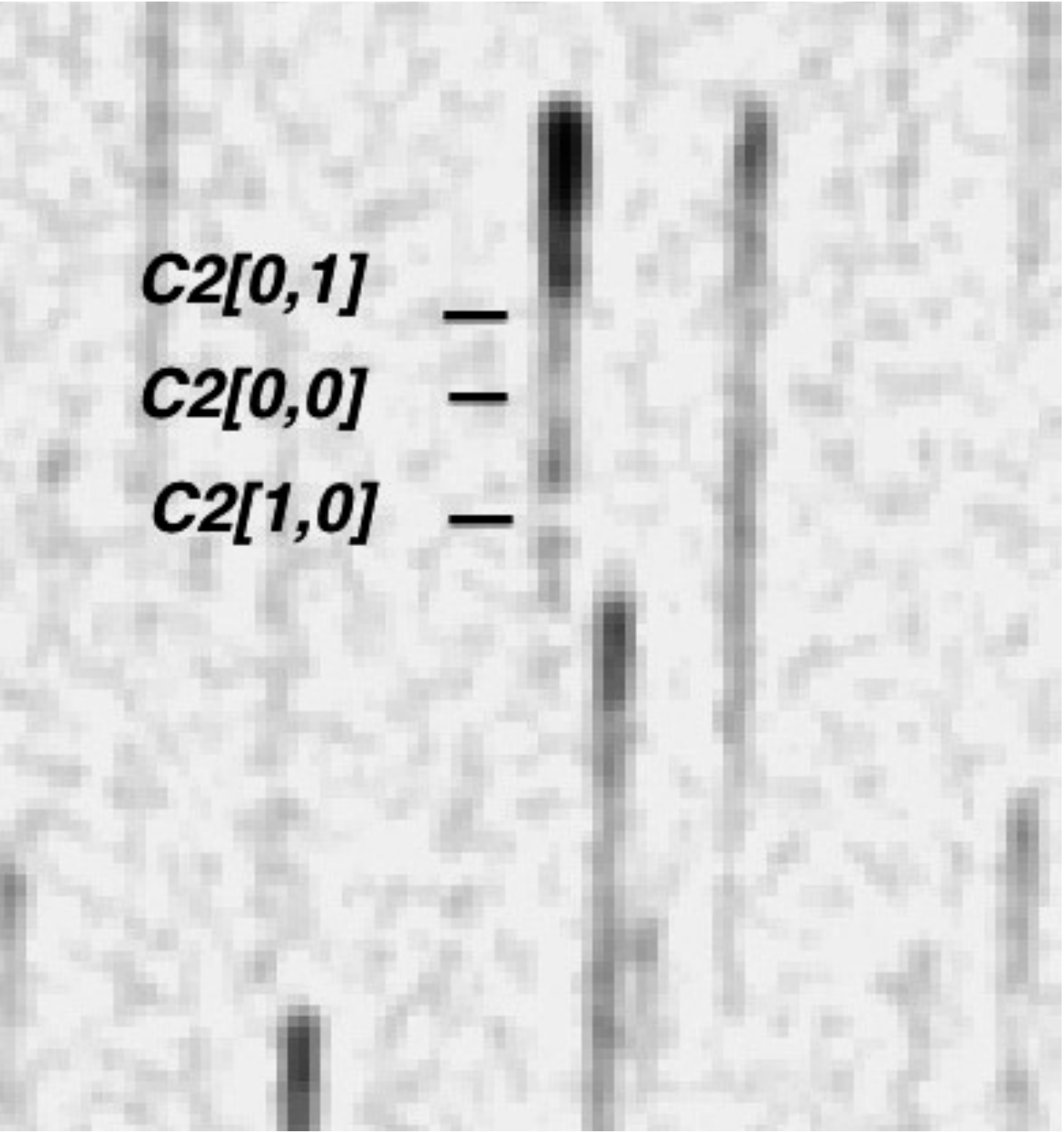}
\includegraphics[width=7cm,height=6cm,angle=0]{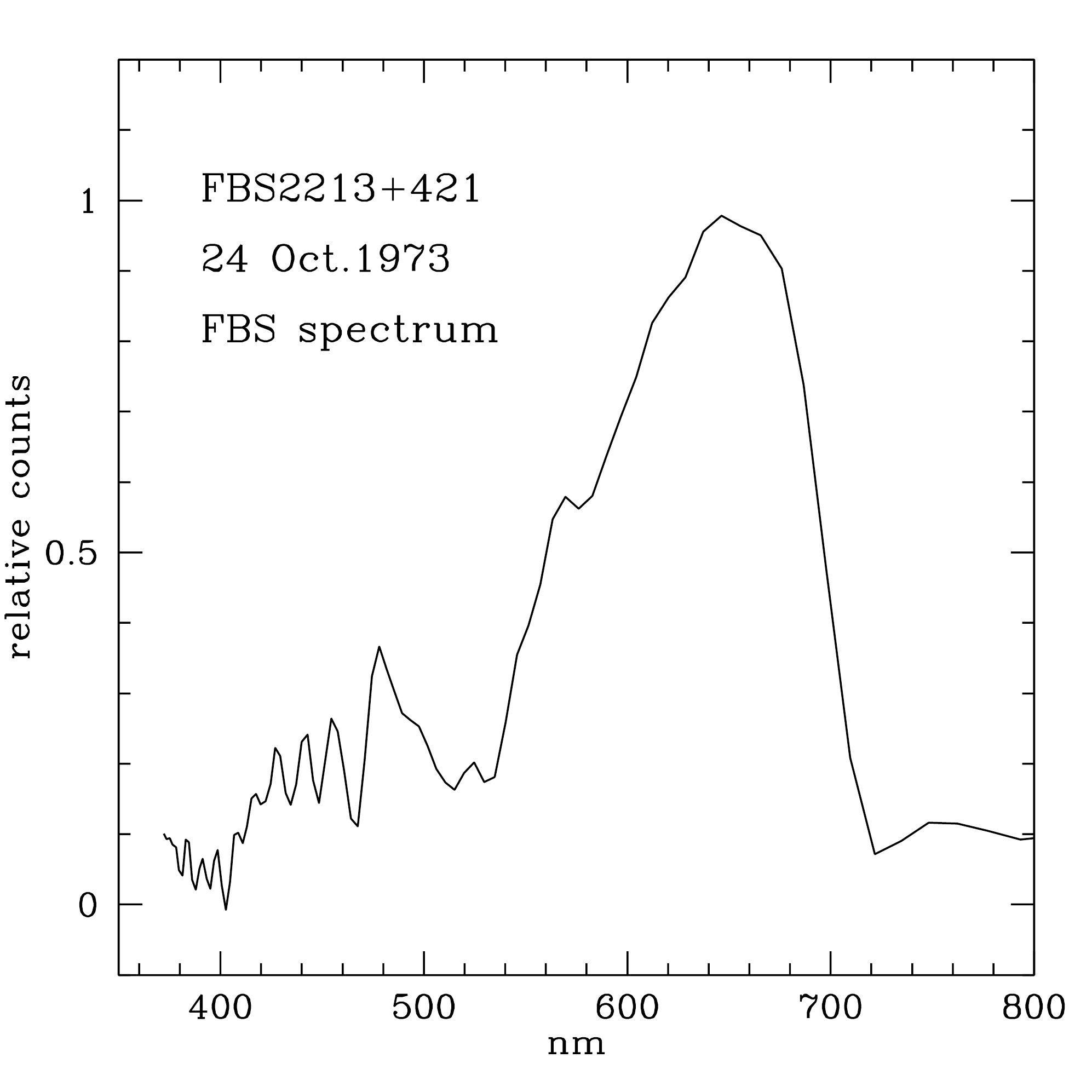}
\caption{ Top:  two dimensional section of the FBS plate No 932. The digitization of V381 Lac shows a strong red head and allowed to identify the deep absorption bands of the  C$_2$ from top to bottom at 5636, 5165,4737\AA, respectively . 
 Bottom: wavelength  calibrated spectrum ;   the central deep  depression is partly due to  the lower sensitivity  of the plate emulsion. }
\label{byuplate}
\end{figure}

\section{ OBSERVATIONAL MATERIAL }    \label{Obs}

\subsection {Optical photometry and spectroscopy} \label {optic}

Between July 2012 and September 2015 we obtained  simultaneous  photometric ($ B,V, R, i$) and spectroscopic data  with the 152cm and 182cm telescopes  of the Bologna and Asiago Astronomical Observatories, equipped with the BFOSC and AFOSC (Bologna/Asiago Faint Object Spectrometer And Camera);  spectral range is 3900-8500\AA,  dispersion 3.9 \AA/pixel and resolution of about 10 \AA . All the spectra were corrected for the atmospheric extinction and normalised at the same wavelength.
One photometric point was also obtained with the Roma University telescope (TACOR) and one was kindly obtained for us by the ARA association with the 37 cm telescope in Frasso Sabino.
The data were reduced by means of standard IRAF \footnote {IRAF is distributed by the NOAO which is operated by AURA under contract with NFS}  procedures. 

To study the photometric history of V381 Lac we have downloaded from MAST   and analysed the digitised images available from POSS\,I, POSS\,II and Quick-V surveys in the $B, V, R$ ~and $I$~ bandpasses.
Well sampled light curves were obtained from the NSVS- ROTSE telescope \citep{woz04a} and the AAVSO  \footnote{American Association of Variable Star Observers, URL: http://www.aavso.org/vstar/vsots/0100.shtml} archive.
 In the  archives of Asiago Observatory we examined the plates of the field, obtained between 1967 and 1969; only an upper limit of $B$=17.7 mag could be derived.   We will not mention these plates anymore.
  We also re-analised the objective prism spectrum  visible in the plate of the Digitized First Byurakan Survey, now accessible from the italian Virtual Observatory  (http://ia2.oats.inaf.it/). 
The  spectrum, showing the deep absorption bands typical of a carbon star, is presented in Figure \ref{byuplate}. 
  From this spectrum we derived  approximate magnitudes in the $B$ ~and $R$ filters  following the same criteria  of the DFBS automatic pipeline \citep {mic07}. 
We summarize the archive and our  recent  optical data in Table \ref {optarch}; typical errors are  $\sim$0.07 mag for the Loiano, Asiago and Frasso Sabino, 0.1 mag for TACOR.    

\begin{table*}
\caption[] {  Archive and new data in the optical range }
\label{optarch}
\begin{flushleft}
\begin{tabular}{ccccccc}
\hline
\noalign{\smallskip}
emulsion/ & date  &JD  & $B$ & $V$ & $R$ & $I$   \\
instrument &   &  &  &  &  &    \\
  \hline
  103aO  & 26 Aug. 1952 & 2434250 & 17.07  (0.14) &   &             &   \\
  103aE  &  26 Aug. 1952 & 2434250 &  &    &  13.34 (0.13) &  \\
  103aF  & 24 Oct. 1973 & 2441979 & 15.6 0.2 &  & 13.3   (0.2) &   \\
  IIaD &   30 Sep. 1983 &2445607   &   & 16.86  (0.10) &  &  \\
 IIIaF   & 4 Sep. 1989 & 2447773 &  &  &  20:: &  \\
 IIIaF  &  4 Oct. 1989 & 2447803 &  &  &  20:: &  \\
 IV-N    &  6 Oct. 1989 & 2447805 &  &  &  &  $>$20 \\
 IIIaJ  & 29 Sep. 1992 &2 448894  &  21:: &  & &  \\
 IV-N   & 20 Sep. 1995 &  2449980 &  &  &  & 15.65 (0.12)  \\
 IIIaJ  &  24 Oct. 1995 & 2445014 & 18.5  (0.3) &  &  &  \\
 NSVS  & May99-Feb00 & 2451~299-575 &  & &  12.2-14.3$^1$ &  \\ 
  SDSS  & 24 May 2006 & 2453879 & und & und  &und  &   $>$21.3$^2$ \\
 AAVSO  & July-Nov. 2012 & 2456~109-262 & 17.5-15.8 & 15.5-14.0 &  & \\
  & &  &    &  &  & \\
BFOSC &18 Jul. 2012 &   2456126   & 17.50  & 14.95   & 13.25 & \\
TACOR & 14 Oct. 2012 &   2456214  &             & 13.7  & 12.4 &   \\
F.Sabino & 18 Oct. 2012 & 2456218   & 16.09   & 13.75  & 12.33  &  \\
AFOSC & 20 Nov. 2012  &  2456251  &  16.20   & 13.67  & 12.20 &  \\
BFOSC  & 28 Nov. 2013  & 2456624  &              &17.40  & 15.61 & \\
BFOSC & 17 Dec. 2013  &  2456643  & 19.70    & 17.65 & 15.93  & 14.21  \\
BFOSC & 21 Aug. 2014 &   2456890  &  $>$ 21.5 & 19.5  & 17.85  &   \\
BFOSC & 13 Sep. 2014 &   2456913  &                 &          & 18.17  &   \\
BFOSC & 20 Jun. 2015 &   2457194  &                 &  17.5   & 15.45  & 14.32  \\
BFOSC & 22 Jul. 2015 &   2457225   &  20.2        &  17.81 & 15.75  & 14.61  \\
AFOSC & 06 Sep 2015 &   2457272  &                 &  19.99  & 18.50$^3$ & 16.87$^3$ \\
\hline
 \noalign{\smallskip}
\multicolumn{5}{l}{ Notes: $^1$ ROTSE Red magnitudes; $^2$ SDSS  magnitude; $^3$  converted to Johnson-Cousin system.  }
\end{tabular}
\end{flushleft}
\end{table*}

%
%
\begin{figure*}
\centering
\includegraphics[width=13cm,height=10cm,angle=0]{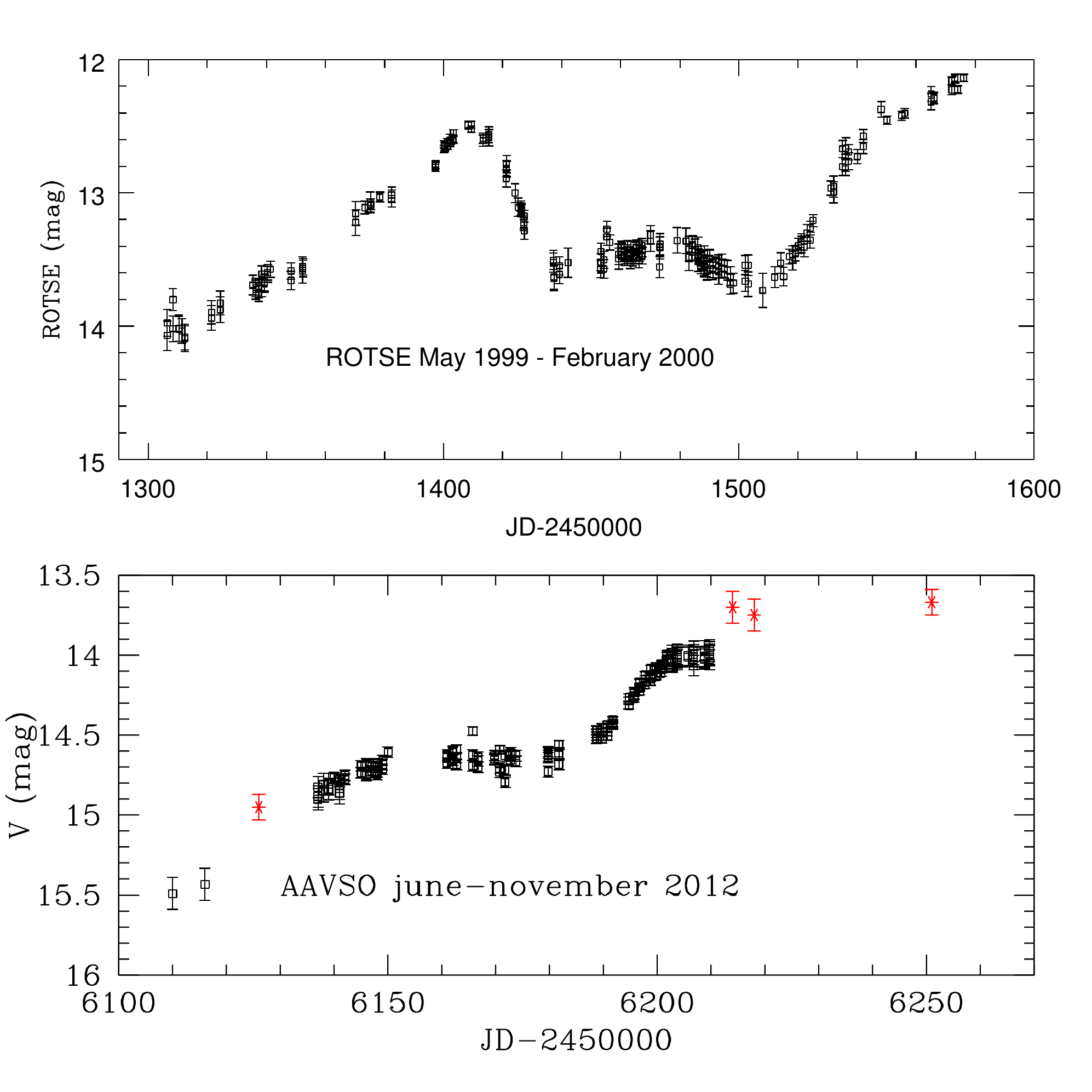}
\caption{ Light curve of FBS2213+421 downloaded from ROTSE and AAVSO archives. The asterisks represent our  2012  data obtained  in the same period of the AAVSO data; a small  offset between  AAVSO V -prevalidated  data  and our filters is evident.  }
\label{rotseaavso}
 \end{figure*}
%

%
\begin{figure*}
\centering
\includegraphics[width=14cm,height=12cm,angle=0]{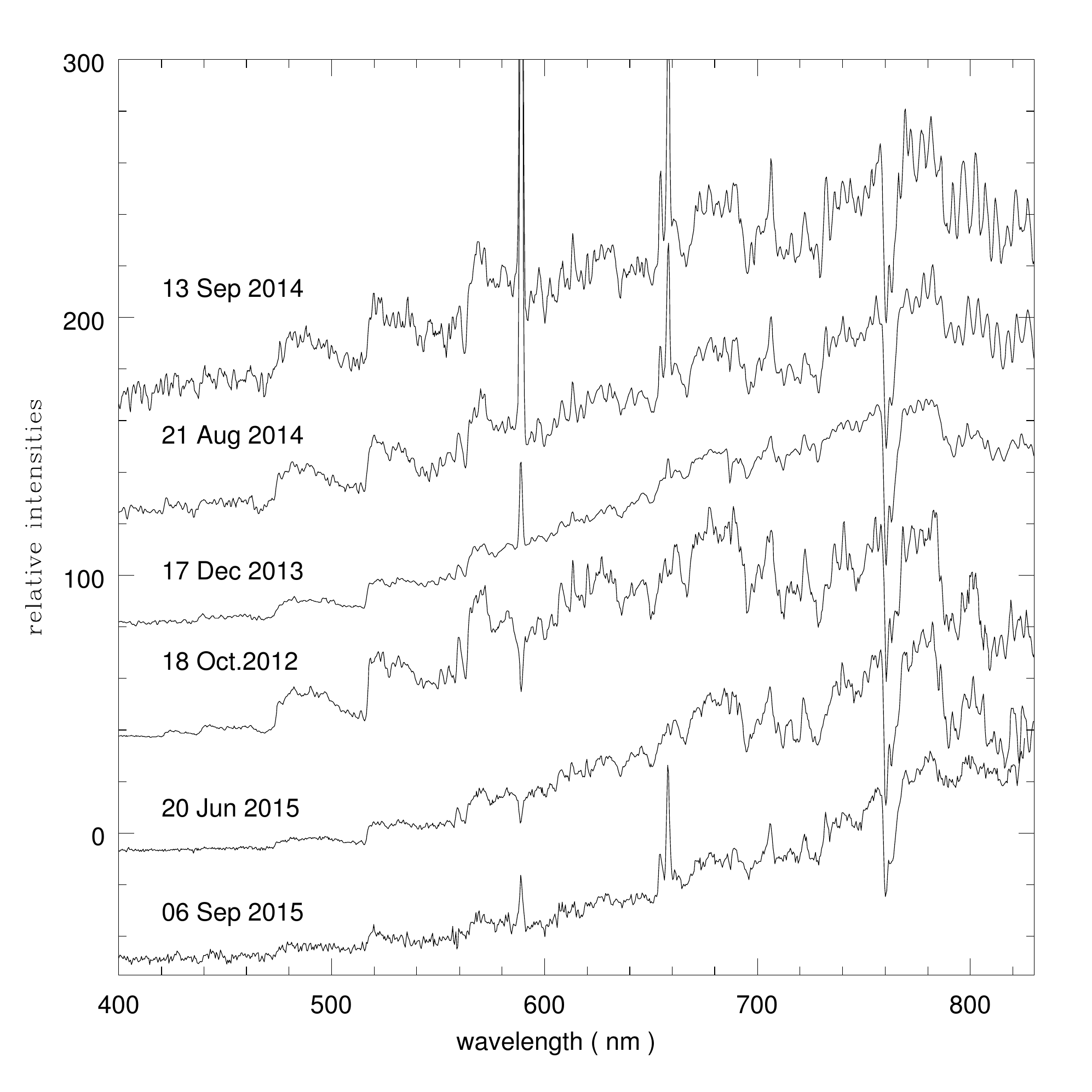}
\caption{ 
 Optical spectra of V381 Lac taken  between 2012 and 2015.  The $y$ axis represents relative intensities corrected for the atmospheric extinction. The spectra are normalise at 7800 \AA. To allow for a better reading the 2013 and 2014 spectra are shifted upward with respect to the 2012 one; the 2015 spectra are placed at the bottom of the figure.}
\label{spectra}
 \end{figure*}

\medskip

\subsection {  Infrared Photometry} \label{irdata}

We obtained  photometric observations in $J$, $H$, and $K$ bands at the 1.1m AZT-24 telescope located at Campo Imperatore (Italy) equipped with the imager/spectrometer SWIRCAM \citep{dalessio}, based on a 256x256 HgCdTe PICNIC array. 
From the archives we have also retrieved the  data from 2MASS \citep{skru2006},
IRAS, WISE \citep{cutri2013}, AKARI \citep{murakami2007} 
 and  converted fluxes to magnitudes according to the prescriptions of \citet{tan09} and of  the \cite{irasexp88}  for  AKARI and IRAS, respectively.
 Recent and archive magnitudes are reported in Table \ref {irarch}. 
We do not report the very uncertain value of  the WISE 4.6$\mu$ magnitude.

\begin{table*}
\caption[] { Recent  and  archive magnitudes in the infrared  range.   }
\label{irarch}
\begin{flushleft}
\begin{tabular}{ccccccccccc}
\hline
\noalign{\smallskip}
  \hline
   source  & date & 1.25 & 1.65 & 2.26 & 3.4  &  9.0   &  12 &  18 &  22/25 &  60    \\
     \hline    
   2MASS (1) & 11-10-1998  & 13.207 & 10.196  &  7.788 &  & & & &  &  \\
    C.I. (2)       &  21-09-2012 &  8.96   & 7.45       &  6.15  & &  & & &  &  \\
    C.I. (3)       & 17-12-2013  & 10.77  & 8.47      &  6.54   &  & & & &  & \\
    C.I. (4)       & 29-09-2014  & 12.74  & 9.65      &  7.33   &  & & & &  & \\
   C.I. (5)        & 22-06-2015  & 11.13  & 9.33      &  7.40   &  & & & &  & \\
 C.I. (6)          & 23-07-2015  & 11.37  & 9.4        &  7.6     &  & & & &  & \\
  IRAS            & 1983           &      &  &  &   &   & 0.626  &   & -0.17  & -0.45  \\
  AKARI         &  2006-2007  & & &  &       &    1.13 &  & -0.14   &  &        \\
    WISE        & 18-06-2010   & & &  & 4.438   &    & 0.365  &    &   -0.227 &   \\     
  \hline
 \noalign{\smallskip}
\multicolumn{8}{l}{Wavelengths are in $\mu$m.}\\
\multicolumn{8}{l}{ C.I. = Campo Imperatore, K is in 2MASS system.  Errors for  July 2015  $H$ ad $K$  are 0.15 mag}
\end{tabular}
\end{flushleft}
\end{table*}

\section {DATA ANALYSIS} \label{DA}

\subsection{Flux variations} \label{flvar}

From the photometric optical history presented in Table \ref{optarch} it is clear that two deep minima occurred in 1989 and 2006. On shorter time scales, variability of $\sim$2 mag are shown by the ROTSE and the AAVSO optical light curve, but no periodicity can be identified (see Figure \ref{rotseaavso}).

In the near infrared our data show large changes with respect to 2MASS, and also within one year, strongly correlated with the optical photometric and spectroscopic variations. At longer wavelengths  the excursions seem to be smaller on the basis of the data presented  in Table \ref{irarch}, obtained at different epochs by different instruments. 

\subsection{Spectroscopic variations}
Our new spectra of V381 Lac at different epochs  are shown in Figure \ref{spectra}. Overall, the spectral changes are strongly correlated with the optical flux and with the sign of its derivative.
  The spectra obtained during  bright phases (1973 and 2012) showed strong absorption bands;  in  the 2012 spectrum  the  bands typical of a naked N type giant, belonging to the Swan system of the C2  and to the red system of CN molecule, could be easily identified.
From a comparison with the spectral atlas of carbon stars by  \citet{barn96} the spectral class of the object was not earlier than N5 subtype. The Na D at 5890-96 \AA ~ doublet was  strong in absorption.
In December  2013, during a fading phase  the continuum was brighter redwards  7800\AA ~and  the veiling started to affect the  strength of the photospheric  absorption bands. 
The most interesting features of this spectrum were the Na D doublet  which  changed  to strong emission and a faint emission at about 6582\AA.
 In summer  2014  the star was even fainter, the emissions were  stronger and the second component of the  [N {\II}]  doublet  became  visible at 6547\,\AA. 
    In June 2015 the star was  in a relatively  bright phase:  only absorption features with sign of veiling were visible in the spectrum, while the evolution of the photometry   from June  to  July indicated a new  brightness decline.
   
 Besides the emission of Na D and  [N {\II}]  during the fading phases, we remark the lack of hydrogen lines either in emission or absorption in all the  new spectra,  while  in 2008  H$\alpha$   was   clearly present in absorption  simultaneously with the Na D doublet  \citep{gig09}.

\subsection {Infrared diagrams} \label{irdiag}
\begin{figure}
\centering
\includegraphics[width=8cm,height=8cm,angle=0]{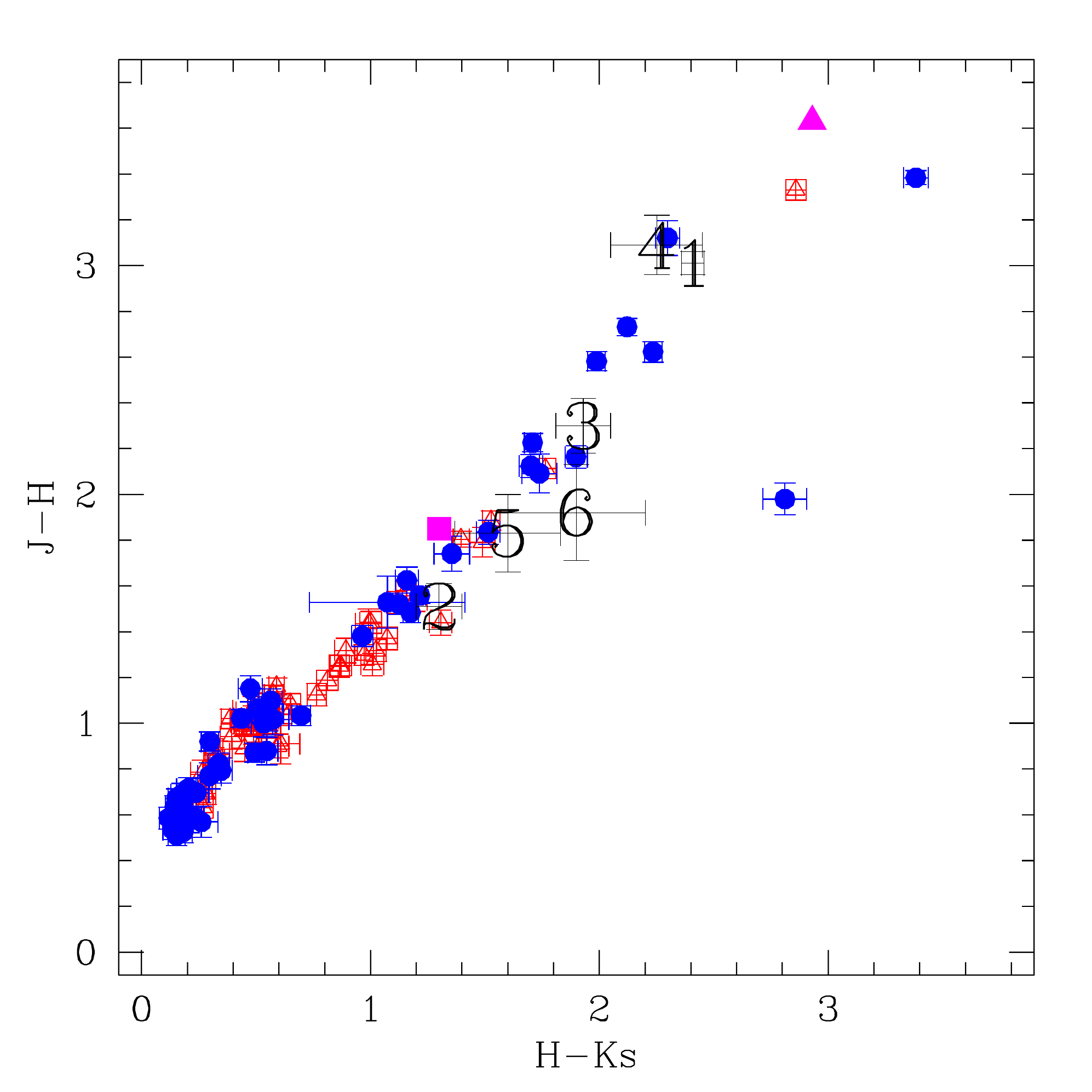}
\caption{  Colour-colour near infrared diagram in the 2MASS system; the positions of V381 Lac are indicated by the black numbers.  Blue filled circles  are the FBS carbon stars, red  open triangles are carbon stars from other dataset.
Theoretical model 1 (magenta square) and model 2 (magenta triangle) are also over imposed.} 
\label{JHK}
\end{figure}

We located  our star in  the classical colour-colour diagram $J-H~ vs~ H-K $  to study its evolution with luminosity \citep[e.g.][]{bessel88, lege02, cruz03}.
We show the results in Figure \ref{JHK} together with those of confirmed carbon stars from FBS and other dataset \citep[][]{alksnis01, mauron08, mauron14}.
The positions of V381 Lac corresponding to the dates reported in Table  \ref{irarch} are indicated by the numbers in black.
The very red 2MASS point in 1988 (point 1) is indicative of the presence of a  cool envelope. In 2012, during  a bright phase, V381 Lac was along the strip of the naked carbon stars (point 2)  while one year later, during the optical fading,  it was moving again towards the dusty  stars region (point 3).
In the autumn 2014, when the star was already faint, the colours were again very similar to those of  2MASS (point 4) .  
 The magnitudes of Summer 2015 (points 5 and 6 ) are intermediate between maximum and minimum. 
 Two points  corresponding to the theoretical models described below are also indicated.

%
%

\begin{figure}
\centering
 \includegraphics[width=7cm,height=7cm,angle=0]{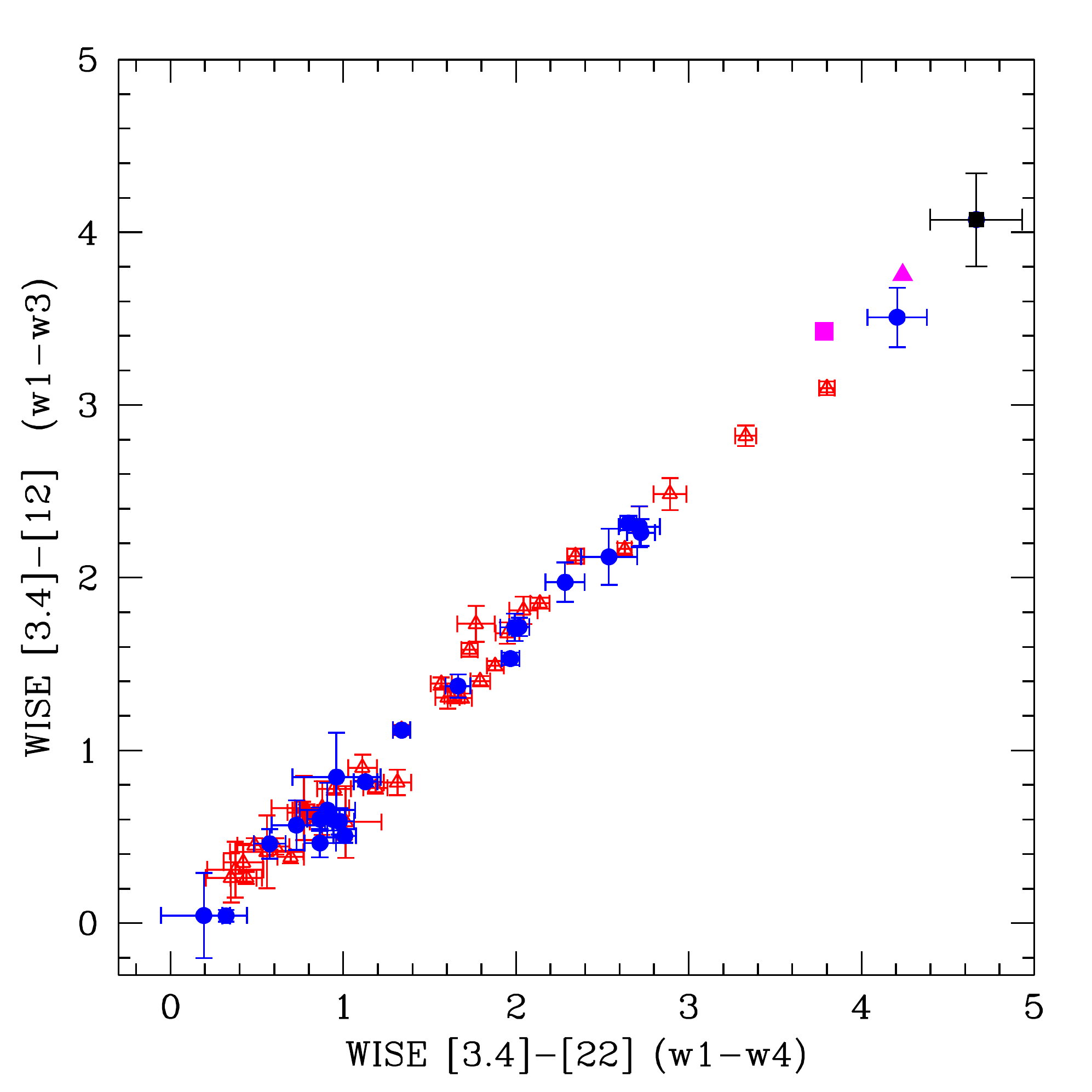}
\caption{  
 WISE w1-w3 vs w1-w4 colour colour diagram for the sample of carbon stars reported in the text. V381 Lac is the uppermost black square. Theoretical model 1 (magenta square) and model 2 (magenta triangle) are also over imposed.}
\label{wise1}
\end{figure}

At longer wavelengths we have checked  the colours of V381 Lac in diagrams involving  the IRAS, WISE  and AKARI magnitudes; some of them are very efficient in discriminating dust enshrouded stars (see fig.8 in \cite{nikutta14},  fig. 4 in \cite{tisse12}, and fig. 3 in \cite{ne14} ). 
In the lack of the  4.6$\mu$~flux in the WISE-2013 database we have  traced the  possible colour - colour diagrams using the three remaining filters. 
An example is given in fig. \ref{wise1}: 
V381 Lac is represented by the red symbol in the region occupied by  obscured carbon stars. 

 We have tried anyway to get more information from WISE, locating V381 Lac in the classical diagram [4.6]$-$[12] vs [12]$-$[22] used by \cite{tisse12}. 
That paper is based on the  WISE Preliminary release \citep{cutri2011} suffering of several problems of calibration. 
We  have applied the appropriate  corrections  to the magnitudes  of V381 Lac according to figure 3 of \cite{tisse12}; 
 when this release is used V381 Lac has the following 
 colours   [4.6]$-$[12] = 2.60  $\pm$0.13, [12]$-$[22] = 0.75$\pm$0.08.
 In  the WISE-2013 release  the magnitude [4.6]  is reported in red (not reliable), the other magnitudes  also changed, while the  colours remain similar;  taking into account the large uncertainties  in the new release too,  we  consider  the above reported  colours  still  acceptable,  that locate the star in the zone populated by known RCB stars  in  figure 6 of \cite{tisse12}.

%

  Similar indications are  derived from the other possible WISE diagrams and from the diagrams involving AKARI data,  taken when the star was in a faint phase. 
  From the IRAS magnitudes computed according to the definition of  \cite{veen88}   the colours of V381 Lac are:  [12]-[25]=-0.796  and [25]-[60]= -1.60 in  region VII of their figure 5b, populated by stars with evolved carbon rich circumstellar shells, in agreement with the previous results.
                           
\begin{figure}
\centering
\includegraphics[width=8cm,height=8cm,angle=0]{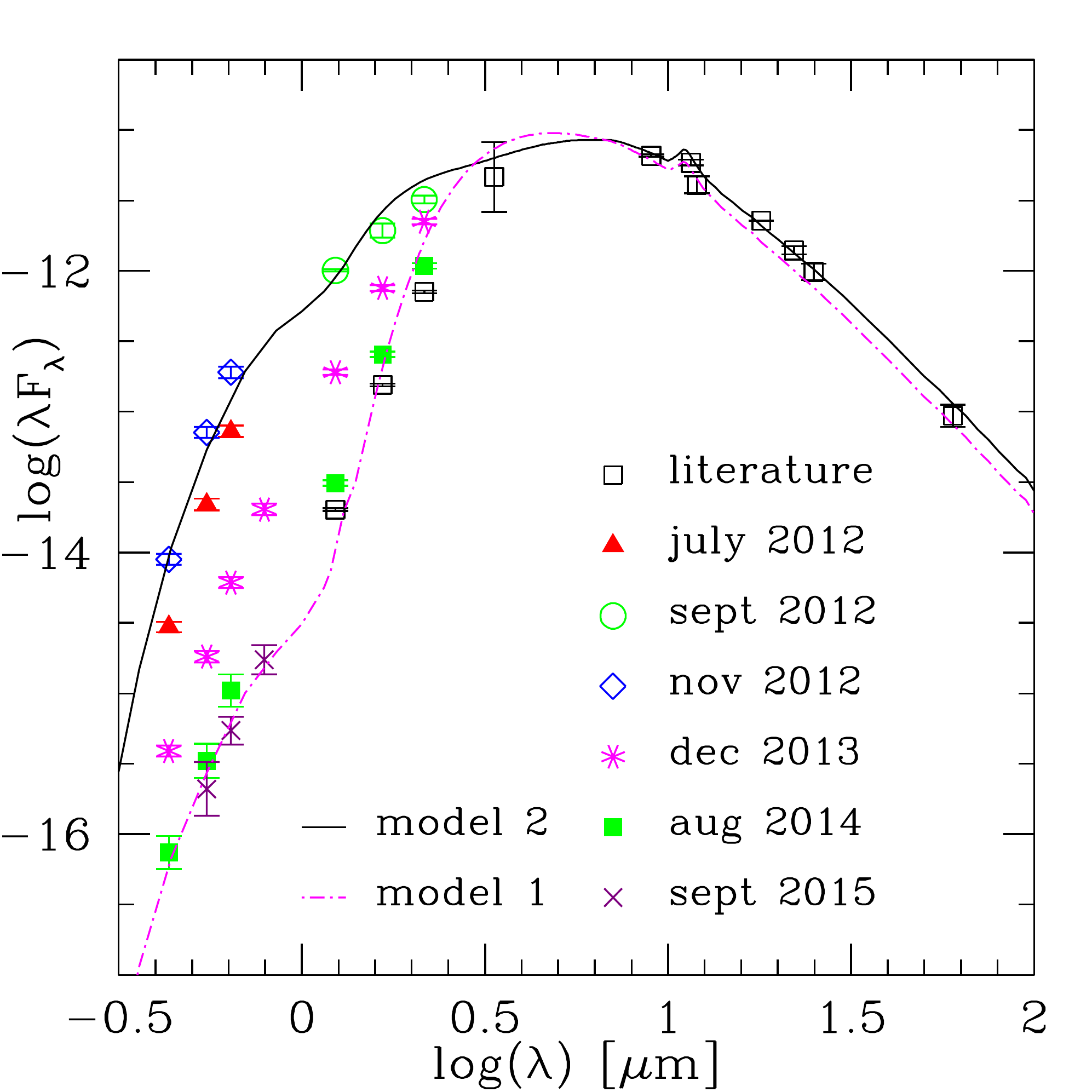}
\caption{Spectral energy distribution of V381 Lac. Fluxes are in SI units. The literature data are  from 2MASS,  AKARI, IRAS, and WISE catalogs. 
The best fit models (see section \ref{theo} for the maximum (magenta dashed line) and minimum (black solid line) obscuration phases are over imposed.}
\label{SEDtot}
\end{figure}

 Finally,  using the archive and our more recent data we have traced the Spectral Energy Distribution for V381 Lac reported  in Fig. \ref{SEDtot}. 
  In the infrared the AKARI data are well aligned with the WISE and IRAS ones, indicating a smaller variability of the source in the mid--infrared.
    To avoid confusion  we do not plot  the  June and July 2015 fluxes. These points overlap quite well those of December 2013 when we caught  the star during a fading  phase.  Although  in summer 2015 no emission  was  visible in the spectrum,  from the comparison of  June and July photometry  we  argued that a  similar evolution was  globally occurring. Our guess was confirmed by  the data  obtained in September.  In Fig. \ref{SEDtot} we include these last points  which correspond to the minimum brightness we have recorded so far.  
 
\section{THEORETICAL MODELS}  \label{theo}

Following the pioneering exploration by \citet{fg06}, various research groups worked on the description of dust formation in the winds of AGBs \citep{nanni13a, paperIV}. In a recent paper, \citet{flavia15} addressed the evolution of the infrared colours of AGB dusty models.
In the present paper we compare the observations of V381 Lac with the predictions of these models, to give a possible theoretical characterization of this star. 

\subsection{The model}

The evolution from the pre--main sequence to the end of the AGB phase was calculated by means of the ATON code for stellar evolution \cite[see][and references therein]{ventura09}. The range of masses investigated is $ 1M_{\odot} \leq M \leq8M_{\odot}$, assuming a chemical composition in the range $0.008\leq Z \leq 0.018$ ($Y=0.26$). The convective instability is treated according to the Full Spectrum of Turbulence (FST) description, developed by \citet{cm91}. For carbon stars, we use the mass loss rate by \citet{wachter08}. 
For low mass models ($1.25\leq M/M_{\odot} \leq 3$), the Third Dredge Up (TDU) drives the bottom of the envelope to regions previously touched by 3$\alpha$ nucleosynthesis; the consequent increase in the surface carbon content eventually leads to the C--star stage. 
This paper is focused on a carbon star and for this reason, we have applied only the models with mass below $3M_{\odot}$.
The wind surrounding AGBs is assumed to be accelerated by radiation pressure on dust particles. The growth of dust grains is described by condensation of gas molecules in the wind impinging on the already formed grains. The relevant equations can be found, e.g., in \citet{fg06}.
For a given AGB evolutionary time, the location (and temperature) of the inner dust region boundary, the density stratification, the relative percentages of the different dust species (i.e., SiC vs. AmC), the optical depth and the dust grain size are calculated self-consistently by our wind AGB models (Ventura et al. 2014 and references therein) and then used as inputs for the DUSTY code  \citep[see][]{flavia15}.


\begin{table}
\caption{Physical properties of the central object and of the circumstellar dust of V381 Lac, obtained by the comparison with models.}
\begin{tabular}{lcc}
  \hline
  \\
  Properties & V381 Lac\\
  \\
  \hline
  Star\\
  $L$ $[L_{\odot}]$  & 8000-10000\\
  $T_{eff} [K]$  &  2500\\
  $M_{MS}$ $[M_{\odot}]$ &  2\\
  $M$ $[M_{\odot}]$  &  1\\
  \\
  Dust\\
 Properties&  model 1 & model 2\\
  T$_{dust} [K]$ & 1000& 600\\
  $\tau_{10}$  &  0.22 & 0.08\\
  $a_{SiC}$ $[\mu m]$  &  0.08 & 0.08 \\
  $a_{C}$ $[\mu m]$  &  0.18 & 0.18\\
  \\
   \hline
\end{tabular}
\label{tabmodel}
\end{table}

\subsection{Comparison and results}

In Fig. \ref{SEDtot}, we can clearly distinguish different phases of obscuration that are particularly evident at optical and near--infrared wavelengths.

Using colors and SED, we identify V381 Lac as the descendant of a star of initial mass $\sim 2M_{\odot}$, in the final AGB phase, after the effects of repeated TDU episodes turned it into a carbon star. Models of smaller mass can be ruled out by the present analysis, because they never reach the observed degree of obscuration.
The gradual increase in the surface carbon favors a considerable expansion of the external regions  \citep{vm09,vm10} and the consequent decrease in the effective temperature, which drops to $\sim2500K$. These conditions favour the formation of great amount of dust, mainly solid carbon, with smaller percentages of SiC.
The SED of the star during the  heavily obscured phase is nicely reproduced (model 2 in Fig. \ref{SEDtot}) by our 2$M_{\odot}$ model, in the evolutionary stage when the mass is reduced to 1$M_{\odot}$.
The dust layer is formed close to the star and has a temperature around 1000 K. Dust is composed by $\sim$80\% of solid carbon and $\sim$20\% of silicon carbide, with grain sizes of $a_C\sim0.18\mu m$ and $a_{SiC}\sim0.1\mu m$ respectively. The optical depth at 10$\mu m$ is $\tau_{10}\sim 0.22$. 
The formation of solid carbon increases the emission in the continuum. The presence of a significant percentage of SiC, evident from the feature at $\sim 11.9 \mu m$ in both models, is confirmed by the slight  increase in the flux of the [12] band with respect to the other WISE bands.

To interpret the 2012 observations, when the SED exhibits a smaller degree of obscuration, we assume that the star experienced a phase of smaller mass loss, during which dust formation was suppressed. In this situation the radiation emitted from the central object is reprocessed by the dust layer formed during a previous phase of stronger mass loss, pushed outwards by effects of radiation pressure. In this configuration (model 1 in Fig. \ref{SEDtot}) the temperature of the dusty layer is $\sim$600$K$, while the composition and dimension of dust grains are preserved  together with the properties of the central object. Also, in this configuration the density decrease determines a lower optical depth, $\tau_{10}\sim 0.08$, with a higher emission in the optical. 
In Table \ref{tabmodel} we summarise the main characteristics of the two different phases described by the models.

Fig. \ref{JHK} and  \ref{wise1} show the same models in the 2MASS and WISE color--color diagrams. 
The reddest colours correspond to the most obscured phase. 

 \section{DISCUSSION AND CONCLUSIONS}   \label{Concl}

Our observations of V381 Lac confirmed its strong photometric variability 
and showed a simultaneous coherent behavior  from  the optical to  the near frared.
A number of  deep minima have been recorded in the last 60 years, most notably in 1989 (POSS\,II) and in 2006 (SDSS) 
Although the historic light curve is rather undersampled no clear sign  of periodicity in the occurrence of fading episodes can be derived.

The magnitude changes  decrease  with increasing  wavelengths.
In the $J-Hvs H-K$ ~diagram the star moves along the strip of the moderately obscured AGB carbon stars, being redder when fainter.
This is  a common behaviour among stars showing erratic variability, and obscuration events. 
\cite{feast97} and  \cite{feast97b}  report an extensive description of different  phenomenologies and several examples of  the evolution in this diagram.
The path followed by the colours of our star is substantially the same in spite of the limited number of our points in $J,\,H,\,K$  filters.

The spectral variations of  V381 Lac  are strongly correlated with the optical flux, showing a behaviour reminiscent of  RCB stars. 
 At maximum, continuum and spectral features are those of a typical N carbon giant. During decline, the photospheric absorptions are progressively diluted by the veiling due to
repeated scattering of the light crossing the soot clouds that eclipse the photosphere.
Na\,D and [N{\II}] doublets go in emission, with increasing intensity as stellar luminosity decreases.
These lines, formed in the circumstellar envelope, are likely present with constant flux at all times, but are not visible during maxima, overwhelmed by the bright central object. Extensive descriptions and explanations of these phenomena may be found e.g. in \cite{clayton96}, \cite{kames2006} and references therein.
 Notably, during a phase of minimum luminosity, corresponding to the maximum emission of Na\,D and [N{\II}] lines, the molecular absorption bands were strong again (see  Fig. \ref{spectra}).
a similar behaviour is reported by \cite{lloy1997} in the framework of  the RCB model  {\it  ' ... as the cloud gradually dissipates the photospheric spectrum returns to normal long before the re-attainment of maximum light' } . 

 In the 2008 spectrum H$\alpha$ was clearly present in absorption;
 the spectrum appears veiled but the photometric phase at that time is unknown.\\
  In the more recent spectra  hydrogen lines have never been observed neither in emission, nor in absorption, while, in similar stars, hydrogen is sometimes observed in emission during dimming. 
  V381\,Lac shows a high level of activity, this suggest a high frequency of gas ejection from the central star,   which can be responsible for a wide variety of variability  between successive events, similarly to what happens e.g. in DY Per, \citep{alksn02}. \\
  The 2008 features  still remain an open question: only a frequent monitoring with simultaneous photometric and spectroscopic observations can help in clarifying this difficult scenario.

Based on the fit of our theoretical models with the SED, V 381 Lac is a carbon star with an initial mass $\sim2 M_{\odot}$ during the last pulses of the AGB phase. The central object with a luminosity around $8\times10^3\leq L/L_{\odot} \leq 10^4$ and effective temperature $T_{eff}\sim 2500K$, is loosing the external envelope with a high rate ($\dot{M} \sim 1.5 \times 10^{-4} M_{\odot}/yr$) producing a dense  carbon-enriched wind. This condition favours the formation of SiC and amorphous carbon in the circumstellar envelope with a grain size of $\sim 0.08 \mu m$ and $\sim 0.18 \mu m$ respectively. The dusty layers enshroud the star, obscuring the stellar radiation in the optical bands. Phases of lower mass loss rate are followed by a decrease in the dust production and a consequent increase of the optical emission.

 We cannot exclude the possible presence of a companion of V381 Lac,  capable to trigger  strong mass loss events 
 like in the case of V Hya, where  the obscuration phases do not occur with regular timing \citep{olivier01}. We still do not have enough observing  material to settle this issue:  again, a long term spectroscopic ( at high resolution ) and photometric monitoring  would be necessary to this purpose.

\section*{ ACKNOWLEDGEMENTS}
We  thank the staff of the  Asiago and Loiano observatories for the allocated time and excellent support for the  observations. 
This research has made use of the SIMBAD database, operated at CDS, Strasbourg, 
France;
~~the Two Micron All$-$Sky Survey database, which is a joint 
project of the University Massachusetts and the Infrared Processing and 
Analysis Center/California Institute of Technology;
~~the data from the Wide$-$field Infrared Survey Explorer, which is a joint project of the University of 
California, Los Angeles, and the Jet Propulsion Laboratory/California 
Institute of Technology;
 ~~the Northern Sky  Variability Survey (NSVS) created jointly by the Los Alamos National 
Laboratory and University of Michigan;
 ~~the AAVSO International Database contributed by observers worldwide. 
~~We finally  wish to thank our referee G.Clayton for his  valuable suggestions which contributed to improve  the  quality of the paper.

\end{document}